\documentclass[aps,prl,reprint,groupedaddress]{revtex4-1}
\usepackage{amssymb}
\usepackage{color}
\usepackage{graphicx}
\usepackage{epsfig}
\usepackage{epstopdf}

\newcommand{\grl}{    {Geophys. Res. Lett.}}
\newcommand{\jgr}{    {J. Geophys. Res.}}
\newcommand{\ssr}{    {Space Sci. Rev.}}
\newcommand{\planss}{    {Plan. Sp. Sci.}}

\newcommand{\solphys}{ {Solar Physics}}

\newcommand{\aapr}{    {The Astronomy and Astrophysics Review}}
\newcommand{\mnras}{    {Monthly Notices of the Royal Astronomical Society}}

\newcommand{\Ham}{\mathcal{H}}

\newcommand{\M}{\mathcal{M}}
\newcommand{\s}{\mathcal{S}}
\newcommand{\Q}{\mathcal{Q}}

\begin{document}

% Use the \preprint command to place your local institutional report
% number in the upper righthand corner of the title page in preprint mode.
% Multiple \preprint commands are allowed.
% Use the 'preprintnumbers' class option to override journal defaults
% to display numbers if necessary
%\preprint{}

%Title of paper
\title{On a transitional regime of electron resonant interaction with whistler-mode waves  in inhomogeneous space plasma}

% repeat the \author .. \affiliation  etc. as needed
% \email, \thanks, \homepage, \altaffiliation all apply to the current
% author. Explanatory text should go in the []'s, actual e-mail
% address or url should go in the {}'s for \email and \homepage.
% Please use the appropriate macro foreach each type of information

% \affiliation command applies to all authors since the last
% \affiliation command. The \affiliation command should follow the
% other information
% \affiliation can be followed by \email, \homepage, \thanks as well.
\author{Artemyev A.V.$^{1,2}$}
\email[]{aartemyev@igpp.ucla.edu}
%\homepage[]{Your web page}
%\thanks{}
%\altaffiliation{}

\author{Neishtadt A.I.$^{2, 3}$}
%\email[]{Your e-mail address}
%\homepage[]{Your web page}
%\thanks{}
%Collaboration name if desired (requires use of superscriptaddress
%option in \documentclass). \noaffiliation is required (may also be
%used with the \author command).
%\collaboration can be followed by \email, \homepage, \thanks as well.
%\collaboration{}
%\noaffiliation

\author{Vasiliev A.A.$^2$}

\author{Mourenas D.$^{4}$}

%\author{Mourenas D.$^4$}
%\email[]{Your e-mail address}
%\homepage[]{Your web page}
%\thanks{}
%\altaffiliation{}
\affiliation{$^1$ Institute of Geophysics and Planetary Physics, University of California, Los Angeles, CA, USA; $^2$ Space Research Institute RAS, Moscow, Russia; $^3$ Department of Mathematical Sciences, Loughborough University, Loughborough LE11 3TU, UK;\\$^4$ LPC2E, CNRS, Orleans, France}%; $^4$ CEA, DAM, DIF, Arpajon, France}
%\affiliation{Department of Mathematical Sciences, Loughborough University, Loughborough LE11 3TU, UK.}

%Collaboration name if desired (requires use of superscriptaddress
%option in \documentclass). \noaffiliation is required (may also be
%used with the \author command).
%\collaboration can be followed by \email, \homepage, \thanks as well.
%\collaboration{}
%\noaffiliation

\date{\today}

\begin{abstract}
Resonances with electromagnetic whistler-mode waves are the primary driver for the formation and dynamics of energetic electron fluxes in various space plasma systems, including shock waves and planetary radiation belts. The basic and most elaborated theoretical framework for the description of the integral effect of multiple resonant interactions is the quasi-linear theory, that operates through electron diffusion in velocity space. The quasi-linear diffusion rate scales linearly  with the wave intensity, $D_{QL}\sim B_w^2$, which should be small enough to satisfy the applicability criteria of this theory. Spacecraft measurements, however, often detect whistle-mode waves sufficiently intense to resonate with electrons nonlinearly. Such nonlinear resonant interactions imply effects of phase trapping and phase bunching, which may quickly change the electron fluxes in a non-diffusive manner. Both regimes of electron resonant interactions (diffusive and nonlinear) are well studied, but there is no theory quantifying the transition between these two regimes. In this paper we describe the integral effect of nonlinear electron interactions with whistler-mode waves in terms of the time-scale of electron distribution relaxation, $\sim 1/D_{NL}$. We determine the scaling of $D_{NL}$ with wave intensity $B_w^2$ and other main wave characteristics, such as wave-packet size. The comparison of $D_{QL}$ and $D_{NL}$ provides the range of wave intensity and wave-packet sizes where the electron distribution evolves at the same rates for the diffusive and nonlinear resonant regimes. The obtained results are discussed in the context of energetic electron dynamics in the Earth’s radiation belt.
\end{abstract}

% insert suggested PACS numbers in braces on next lines
% insert suggested keywords - APS authors don't need to do this
%\keywords{}

%\maketitle must follow title, authors, abstract, \pacs, and \keywords
\maketitle
\section{Introduction}
Wave-particle resonant interactions in space plasma play a crucial role in momentum and energy exchange between different charged particle populations in absence of particle collisions. Two main regimes of such interactions are quasi-linear diffusion \citep{Vedenov62,Drummond&Pines62} and nonlinear resonant interactions \citep{ONeil65,Mazitov65}. The most advanced theoretical approaches for these two regimes \citep{Kennel&Engelmann66, Karpman74:ssr} have been developed to model the dynamics of energetic electron populations in the Earth magnetosphere, where electrons interact resonantly with whistler-mode waves. Electron interactions with low-amplitude broadband whistler-mode waves are well described by the diffusion model proposed by \citep{Kennel&Petschek66}, whereas the model of resonant interactions with intense coherent waves describes such fast non-diffusive effects as phase trapping \cite{Nunn71,Karpman74}. Many aspects of these two models have been confirmed and verified with spacecraft observations of long-term (diffusive) electron flux evolution \citep[e.g.,][]{Thorne13:nature,Li14:storm} or rapid (non-diffusive) electron acceleration \citep[e.g.,][]{Agapitov15:grl:acceleration,FosterGRL14,Gan20:grl:II} and losses \citep[e.g.,][]{Mourenas16}. The same two model approaches (quasi-linear diffusion and nonlinear resonances) can describe various wave-particle interactions in the solar wind \citep[e.g.,][]{Voshchepynets17, Kuzichev19, Shaaban19}, planetary and interplanetary shock waves \citep[e.g.,][]{Veltri&Zimbardo93, Bykov&Treumann11}, aurora acceleration region \citep[e.g.,][]{Kletzing94, Shen&Knudsen20}. 

The basic equations for the quasi-linear diffusion model have been initially derived for a homogeneous plasma \citep{Andronov&Trakhtengerts64, Kennel&Engelmann66}, and then generalized for inhomogeneous magnetic traps, like the Earth dipole field \citep{bookLyons&Williams, bookTrakhtengerts&Rycroft08}. An interesting and important aspect of such a generalization is the significant relaxation of the broad wave spectrum requirement \citep[see][]{Karpman74:ssr, LeQueau&Roux87}. Indeed, for coherent waves the resonance width, a crucial parameter for the diffusive model, can be determined by the magnetic field inhomogeneity \citep{Karpman&Shkliar77,Shklyar81, Albert93} and/or the wave frequency drift \cite{Demekhov06,Demekhov09,Omura07}. Electron resonant interactions with  coherent whistler-mode waves in the Earth's magnetosphere can be described by a quasi-linear diffusion model \citep{Albert10} if wave intensity is sufficiently small. Therefore, both regimes of wave-particle interaction (quasi-linear diffusion and nonlinear resonances) can operate for the same coherent monochromatic wave, and only wave intensity determines the relevant regime. Thus, the question arises of the transition between these two regimes \citep[see, e.g., discussion in][]{Tao13,Allanson20,Mourenas18:jgr}.

The probabilistic distribution of whistler-mode wave intensities in the Earth’s magnetosphere contains a large population of low-intensity waves (presumably interacting with electrons diffusively), but this distribution is not Gaussian and there is a significant population of whistler-mode waves sufficiently intense to resonate with electrons nonlinearly \citep{Cully08,Artemyev16:ssr,Tyler19}. The quasi-linear diffusion model relies on the average wave intensity presumably dominated by low-intensity waves, whereas the nonlinear interaction model invokes the much more rarely observed intense waves. How do such rare nonlinear resonances alter the generally diffusive electron flux evolution? And can the contribution of intense waves be described by a simple increase of diffusion rates of the quasi-linear model? To address these questions, we should describe the transition between these two regimes of wave-particle interactions. This topic is the focus of our study. We start with basic equations of wave-particle resonant interactions for the most widespread case of field-aligned whistler-mode waves. Then we derive the main characteristics of resonant electron dynamics for both regimes of resonant interactions. We use the mapping technique for such interactions \citep{Benkadda96,Artemyev20:pop} to derive the main scaling laws of the transition between quasi-linear and nonlinear interactions. Then we discuss the obtained results and summarize them.
 
\section{Field-aligned whistler-mode waves \label{sec:basic}}
We start with the Hamiltonian of a relativistic electron (the charge is $-e$, the rest mass is $m_e$, energy is $E=m_ec^2(\gamma-1)$ where $c$ is the speed of light and $\gamma$ is the Lorentz factor) moving in the inhomogeneous background magnetic field $B_0(s)$ (where $s$ is the field-aligned coordinate) and whistler-mode wave field \citep{Albert93,Vainchtein18:jgr}:
\begin{eqnarray}
 H &=& m_e c^2 \gamma  + \sqrt {\frac{{2I_x \Omega _0 }}{{m_e c^2 }}} \frac{{eB_w }}{{k\gamma }}\sin \left( {\phi  + \psi } \right) \nonumber\\ 
 \gamma  &=& \sqrt {1 + \left( {\frac{{p_\parallel  }}{{m_e c}}} \right)^2  + \frac{{2I_x \Omega _0 }}{{m_e c^2 }}}  \label{eq:01}
 \end{eqnarray}
where $p_\parallel$ is the field-aligned momentum (conjugate to $s$), $\Omega_0=eB_0(s)/m_ec$ is the electron gyrofrequency, $I_x$ and $\psi$ are magnetic moment (normalized in such a way that $2I_x\Omega_0 $ is the perpendicular energy)  and the conjugate gyrophase, $B_w(s)$ is the wave magnetic field amplitude, $\phi$ is the wave phase, with $\partial \phi/\partial s=k(s)$ the wave number and $-\partial \phi/\partial t=\omega={\rm const}$ the wave frequency. We use the cold plasma dispersion relation for $kc/\omega_{pe}=(\Omega_0(s)/\omega-1)^{-1/2}$ with a constant plasma frequency $\omega_{pe}$. The spatial scale of the background magnetic field inhomogeneity $R$ is much larger than the wavelength ($Rk\gg 1$), whereas the wave amplitude is much smaller than the background magnetic field ($B_w/B_0\ll 1$). 

Hamiltonian equations for (\ref{eq:01}) are
\begin{eqnarray}
\dot p_\parallel   &=&  - \frac{{I_x }}{\gamma }\frac{{\partial \Omega _0 }}{{\partial s}} - \sqrt {\frac{{2I_x \Omega _0 }}{{m_e c^2 }}} \frac{{eB_w }}{\gamma }\cos \left( {\phi  + \psi } \right), \nonumber \\ 
 \dot s &=& \frac{{p_\parallel  }}{{m_e \gamma }},\quad \dot \psi  = \frac{{\Omega _0 }}{\gamma } + \sqrt {\frac{{2I_x \Omega _0 }}{{m_e c^2 }}} \frac{{eB_w }}{{2I_x k\gamma }}\sin \left( {\phi  + \psi } \right), \nonumber\\ 
 \dot I_x  &=&  - \sqrt {\frac{{2I_x \Omega _0 }}{{m_e c^2 }}} \frac{{eB_w }}{{k\gamma }}\cos \left( {\phi  + \psi } \right) \label{eq:02}
 \end{eqnarray}
where the rates of variation of $I_x$ and $p_\parallel$ are much smaller than the rate of variation of $\psi$, because $\Omega_0 \gg c/R$ and $B_0\gg B_w$. Therefore, $p_\parallel$, $s$, $I_x$ are slow variables, and both the wave phase $\phi$ and gyrophase $\psi$ are fast variables. 

In the absence of a wave ($B_w=0$) the electron moves with constant energy $\gamma$ and magnetic moment $I_x$ (because the Hamiltonian does not depend on $\psi$ and wave phase $\phi(s,t)$). Therefore, each trajectory can be characterized by a pair of initial $(\gamma, I_x)$. Instead of $I_x$, it is often more convenient to use the electron pitch-angle $\alpha_{0}$ determined at the $B_0$ minimum (at the equator, $s=0$): $2I_x\Omega_0(0)/m_ec^2=(\gamma^2-1)\sin^2\alpha_{0}$. The classical magnetic field configuration of the Earth magnetosphere includes $B_0(s)$ growing away from the equator $s=0$ (e.g., the dipole field is $B_0=B_0(0)\sqrt{1+3\sin^2\lambda}/\cos^6\lambda$ with magnetic latitude $\lambda$ defined as $ds/R=\sqrt{1+3\sin^2\lambda}\cos\lambda d\lambda$). In such a field, electrons are moving within a magnetic trap: $p_\parallel=\pm\sqrt{1-\gamma^2-2I_x\Omega_0(s)/m_ec^2}$ is oscillating between $\pm\sqrt{1-\gamma^2-2I_x\Omega_0(0)/m_ec^2}$ values and reaches zeros at the mirror coordinates $s_{\max,\min}$ defined as $\Omega_0(s_{\max,\min})\sin^2\alpha_{0}/\Omega_0(0)=1$. The first-order cyclotron resonance condition $\dot\phi+\dot\psi=kp_z/m_e\gamma-\omega+\Omega_0/\gamma=0$ determines the position of resonance $s_R$ for given energy and pitch-angle $\alpha_0$ with the resonant $p_z=-\sqrt{1-\gamma^2-\Omega_0(s_R)\sin^2\alpha_0/\Omega_{0}(0)}$. 

\begin{figure*}
\centering
\includegraphics[width=0.9\textwidth]{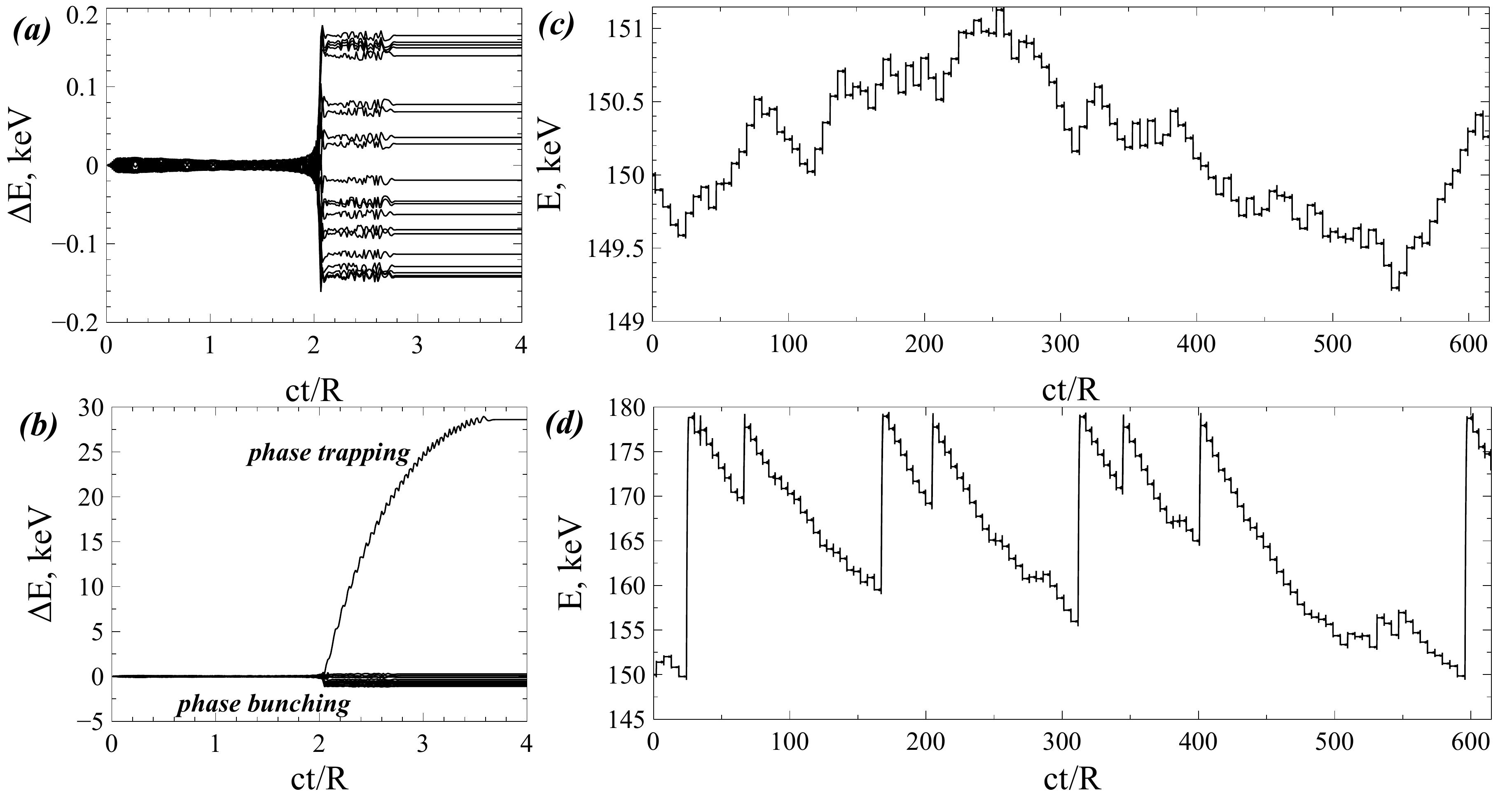}
\caption{
\label{fig1} Trajectories of electrons resonantly interacting with a parallel whistler-mode wave (one resonant interaction is shown): wave amplitude is sufficiently low and electrons are diffusively scattered (a), wave amplitude is high and electrons experience phase bunching and phase trapping (b). Long-term dynamics of electron trajectories ($\sim 100$ resonances) for low (c) and high (d) wave amplitudes. All electrons have the same initial energy of $150$ keV and pitch-angle $45^\circ$. The background magnetic field is the Earth's dipole field with an equatorial magnitude corresponding to $6$ Earth radii away from the planet (the outer radiation belt). Wave characteristics are typical for parallel whistler-mode waves: $\omega/\Omega_0(0)=0.25$, $B_{w}=20$ pT for (a,c) and $B_{w}=200$ pT for (b,d) \citep{Li11, Zhang18:jgr:intensewaves}. The background plasma density is given by an empirical model \citep{Sheeley01}.}
\end{figure*}

The equation for $p_\parallel$ in system (\ref{eq:02}) contains two terms: the mirror force and the wave force. For the diffusive regime of resonant interactions the mirror force is much larger than the wave force, and thus wave effects can be calculated using the integration of wave field over unperturbed electron trajectory. For nonlinear resonant interactions, these two terms should be of the same order, i.e. the wave force should be able to compete with the mirror force and change the electron trajectory substantially. The ratio of the magnitudes of these two forces is $(B_w/B_0)R\Omega_0/c$. Thus, the regime of nonlinear resonant interaction requires $B_w/B_0\sim c/R\Omega_0$, whereas if $B_w/B_0\ll c/R\Omega_0$ the resonant interaction should be diffusive. Figure \ref{fig1} shows electron trajectories for these two regimes: panel (a) shows the energy scattering with a mean zero for $B_w/B_0\ll c/R\Omega_0$, whereas panels (b) shows a significant energy change for $B_w/B_0\sim c/R\Omega$. The nonlinear regime is characterized by two processes: a small number of particles experience phase trapping with a large energy increase, whereas most of the particles experience nonlinear scattering (phase bunching) with a small energy decrease. The long term dynamics (multiple resonances) for a single electron would resemble the random fluctuations of a diffusive regime (see panel (c)) with some sort of cycle for the nonlinear regime (see panel (d)). This cycle consists in multiple drifts to smaller energy due to the nonlinear scattering and rare trappings with energy increase. We aim to describe the transition between these two regimes for an ensemble of electrons. For this reason we shall start with the quantification of the main characteristics of electron dynamics in both regimes. 

\section{Equations of particle motion around the resonance}\label{sec:resonance}
\begin{figure*}
\centering
\includegraphics[width=0.8\textwidth]{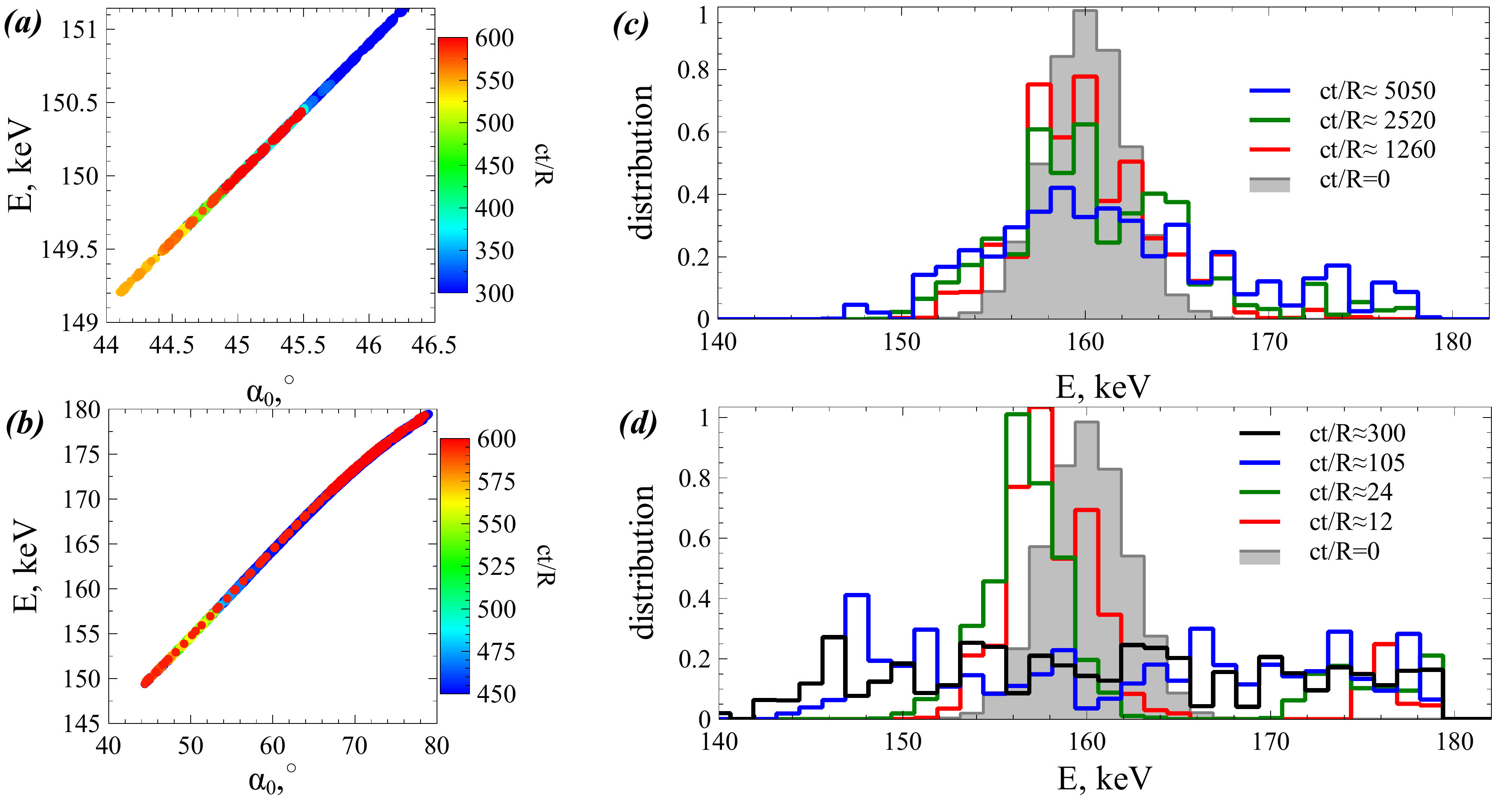}
\caption{
\label{fig2} Two trajectories from Fig. \ref{fig1}(c,d) are shown in (energy, pitch-angle) plane for diffusive (a) and nonlinear (b) resonant interactions (color shows time). Evolution of an initially localized peak of distribution $f(\gamma)$ for the same $h=\gamma-(\omega/2\Omega_0(0))(\gamma^2-1)\sin^2\alpha_0$ as from panels (a,b): diffusive (c) and nonlinear (d) resonant interactions. Note that time-scales are different for (c) and (d) panels.}% \todo{Distributions on (c,d) will be replotted after calculation finish.} }
\end{figure*}

The diffusive regime of resonant interaction can be described by the diffusion coefficient $D_{QL}$, whereas the nonlinear regime is characterized by energy changes due to the nonlinear scattering $\Delta\gamma_{scat}$, energy changes due to trapping $\Delta\gamma_{trap}$, and the probability $\Pi$ of trapping \citep[][]{Albert13:AGU, Shklyar09:review, Artemyev18:cnsns}. In this section we demonstrate the relations between these different characteristics. First, we should consider Hamiltonian (\ref{eq:01}) around the resonance. We follow the procedure described in \citep{Neishtadt75,Itin00} and introduce the new phase $\zeta=\phi+\psi$ through the generating function $W=(\phi+\psi)I+P_\parallel s$. In the new variables, the Hamiltonian takes the form: 
\begin{eqnarray}
 H_I  &=& m_e c^2 \gamma  - \omega I + \sqrt {\frac{{2I\Omega _0 }}{{m_e c^2 }}} \frac{{eB_w }}{{k\gamma }}\sin \zeta  \nonumber\\ 
 \gamma  &=& \sqrt {1 + \left( {\frac{{P_\parallel   + kI}}{{m_e c}}} \right)^2  + \frac{{2I\Omega _0 }}{{m_e c^2 }}}  \label{eq:03} 
 \end{eqnarray}
where we keep the same notation for $s$ conjugate to $P_\parallel=p_\parallel-kI$ and take into account that new Hamiltonian $H_I=H+\partial W/\partial t$. New momentum $I$ conjugate to $\zeta$ equals to $I_x$. Hamiltonian (\ref{eq:03}) does not depend on time, and thus $\gamma-\omega I/m_ec^2= h$ is the integral of motion. This relation between energy and $I$ for a given frequency $\omega$ makes the resonant interaction effectively 1D, i.e. any electron energy and pitch-angle changes obey $\gamma-(\omega/2\Omega_0(0))(\gamma^2-1)\sin^2\alpha_0=h={\rm const}$. Figure \ref{fig2}(a,b) shows such 1D motion in the $(\gamma,\alpha_0)$ plane for diffusive and nonlinear resonant interactions of electrons from Fig. \ref{fig1}(c,d). 

Therefore, we can introduce the 1D electron distribution $f(\gamma)$ for $h={\rm const}$ and follow the evolution of this distribution in these two different regimes. Figure \ref{fig2}(c) shows that the initially localized $f(\gamma)$ is diffusively spread for low wave intensity. Phase trapping and nonlinear scattering modify the initially localized $f(\gamma)$ distribution in a non-diffusive manner: there is a drift of the entire distribution due to nonlinear scattering and the appearance of a high-energy electron population due to trappings (see Fig. \ref{fig2}(d)). An interesting and important property of the wave-particle resonant interaction is that both regimes (diffusion and nonlinear resonances) ultimately result in $f(\gamma)$ relaxation to a plateau with $\partial f/\partial \gamma|_{h={\rm const}}\to 0$ \citep[see][]{Artemyev19:pd}. The energy ranges are different for the two regimes, and the diffusive relaxation would work for a wider energy range because $B_w/B_0\sim c/R\Omega_0$ can be satisfied only for a certain $\gamma\in[\gamma_{\max},\gamma_{\min}]$, where $\gamma_{\max,\min}$ are determined by $B_0(s)$ and $B_w(s)$ profiles. The time-scales of $f(\gamma)$ relaxation are also different for the two regimes: the diffusive relaxation requires more time and Fig. \ref{fig2}(d) shows how $f(\gamma)$ continues spreading outside the energy range of nonlinear resonances where the plateau forms first. 

The resonance condition for Hamiltonian (\ref{eq:03}) is given by equation $\dot\zeta=0$ and this equation determines the resonant momentum $I_R$:
\begin{eqnarray}
 \frac{{kI_R }}{{m_e c}} &=&  - \frac{{P_\parallel }}{{m_e c}} - \frac{{\Omega _0 }}{{kc}} + \left( {\left( {\frac{{kc}}{\omega }} \right)^2  - 1} \right)^{ - 1/2}  \nonumber\\ 
  &\times& \left( {1 - \left( {\frac{{\Omega _0 }}{{kc}}} \right)^2  - 2\frac{{\Omega _0 }}{{kc}}\frac{{P_\parallel }}{{m_e c}}} \right)^{1/2} \label{eq:04}
 \end{eqnarray}
We substitute $I_R$ into the $\gamma$ expression and obtain the resonant energy
\begin{equation}
\gamma _R  = \frac{{kc/\omega }}{{\sqrt {\left( {kc/\omega } \right)^2  - 1} }}\left( {1 - \left( {\frac{{\Omega _0 }}{{kc}}} \right)^2  - 2\frac{{\Omega _0 }}{{kc}}\frac{{P_\parallel}}{{m_e c}}} \right)^{1/2} \label{eq:05}
\end{equation}
Substituting  $P_\parallel$ from $h=\gamma_R-\omega I_R/m_ec^2$ into Eq. (\ref{eq:05}) we can obtain $\gamma_R(s)$ that directly connects the energy of the resonant electron and the resonant $s_R$, i.e. $h={\rm const}$ determines $\gamma_R=\gamma_R(s_R)$.

To determine the electron dynamics (energy change) around the resonance, we expand Hamiltonian (\ref{eq:03}) as
\begin{eqnarray}
 H_I  &=& m_e c^2 \gamma _R  - \omega I_R \nonumber \\ 
  &+& \frac{1}{2}m_e c^2 g\left( {I - I_R } \right)^2  + \sqrt {\frac{{2I_R \Omega _0 }}{{m_e c^2 }}} \frac{{eB_w }}{{k\gamma _R }}\sin \zeta  \label{eq:07}\\
 g &=& \left. {\frac{{\partial ^2 \gamma }}{{\partial I^2 }}} \right|_{I = I_R }  = \frac{{k^2 }}{{m_e^2 c^2 }} \nonumber
 \end{eqnarray}
Such expansion can be done for $2\Omega_0 I/m_ec^2\gg B_w/B_0$, whereas for smaller $I$ an alternative consideration would be required \citep[see][]{Artemyev21:pop, Albert21}. 

We use the generating function $Q=(I-I_R)\zeta+P_\parallel s$ to introduce $P_\zeta=I-I_R$. The new Hamiltonian consists of two parts $ \Lambda  = m_e c^2 \gamma _R  - \omega I_R $ and $H_\zeta$:
\begin{eqnarray}
 H_\zeta   &=& \frac{1}{2}m_e c^2 gP_\zeta ^2  + \left\{ {\Lambda ,I_R } \right\}\zeta  + \sqrt {\frac{{2I_R \Omega _0 }}{{m_e c^2 }}} \frac{{eB_w }}{{k\gamma _R }}\sin \zeta
\label{eq:08}
 \end{eqnarray}
where $\Lambda$ depends on new variables $p=P_\parallel+\zeta\partial I_R/\partial s$,  $q=s-\zeta\partial I_R/\partial P_\parallel$, and we expand $H_I$ over $\zeta\partial I_R/\partial s$, $-\zeta\partial I_R/\partial P_\parallel$ to get the Poisson brackets $\{\cdot,\cdot \}$ \citep[see, e.g.,][]{Artemyev15:pop:probability, Vainchtein18:jgr}. Hamiltonian $H_\zeta$ describes the electron dynamics on the $(\zeta, P_\zeta)$ plane (around the resonance $P_\zeta=0$), and coefficients of this Hamiltonian depend on slowly changing $(p,q)$. Note that the dependence on $(p,q)$ can be rewritten as a dependence on $\gamma_R$, because $(p,q)\approx (P_\parallel, s)$ where $s=s_R$ and $P_\parallel=P_\parallel (s_R)$ due to $\gamma_R-\omega I_R/m_ec^2=h$ conservation. 

\begin{figure*}
\centering
\includegraphics[width=0.9\textwidth]{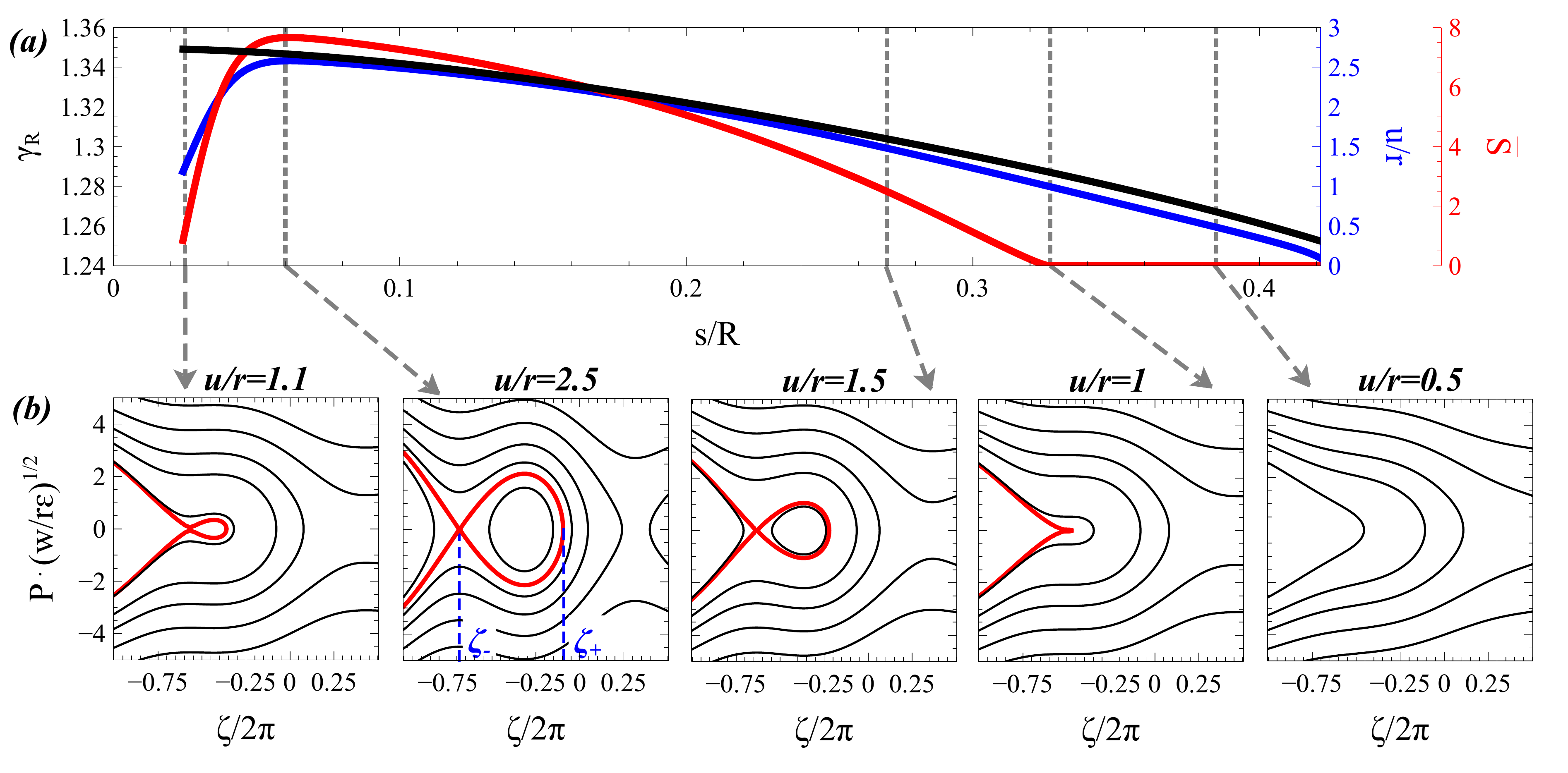}
\caption{
\label{fig3} The profile of $u/r$ ratio (a). Phase portraits of Hamiltonian (\ref{eq:09}) for different $u/r$ ratios (b). Profiles of $\gamma_R$ and $\bar{S}=S\cdot(R\Omega_0(0)/c)^{1/2}$ along $s$ coordinate (note $R\Omega_0(0)/c$ is of the order of $1/\varepsilon$ for the system parameters) (a). System parameters correspond to the trajectory shown in Fig. \ref{fig1}(b)}
\end{figure*}

Hamiltonian (\ref{eq:08}) describes the classical system of a pendulum with a torque \citep[e.g.,][]{bookAKN06}, and the phase portraits of this system with different values of coefficients are shown in Fig. \ref{fig3}(a). To describe the main properties of this Hamiltonian, let us rewrite it in a dimensionless form $\Ham=H/m_ec^2$:
\begin{equation}
\Ham = \frac{1}{2}wP^2  + \varepsilon \left( {r\zeta  + u\sin \zeta } \right) \label{eq:09}
\end{equation}
where $P=P_\zeta/P_0$, $w=gP_0^2$ ($P_0$ is the $P_\zeta$ magnitude), $\max{B}_w/B_0(0)=\varepsilon$, $u=\sqrt{2I_R\Omega_0/m_ec^2}\left(\Omega_0(0)/ck\gamma_R\right)(B_w/\max B_w)$, and $r=\{\Lambda, I_R\}/m_ec^2\varepsilon$, and time renormalized to $1/P_0$ to keep Hamiltonian form of equations. Note $\{\Lambda, I_R\}\propto c/R\Omega_0$ and we assume that $ c/R\Omega_0 \leq B_w/B_0$. 

For a typical wave field distribution along magnetic field lines ($B_w=0$ at the equator $s=0$ and at high magnetic latitudes $s\to 1$, whereas there is a $B_w$ maximum at the medium latitudes \citep[see][]{Agapitov13:jgr, Agapitov18:jgr}), there are three spatial regions with different phase portraits of Hamiltonian (\ref{eq:09}). The wave is generated near the equator $s=0$ and moves to higher $s$ (i.e., wave number $k>0$), whereas the resonant electron moves from high $s$ to the equator because the resonant momentum $p_\parallel=(\gamma\omega-\Omega_0)/k$ is negative for not-too-high $\gamma$. Thus, we consider the phase portrait evolution moving from large $s$ to the equator. At high $s$ wave amplitude is small and $|r|>u$. Thus, the phase portrait totally consists of trajectories crossing the resonance $P=0$ once (open trajectories). As $s$ decreases, we obtain $u>|r|$. The corresponding phase portrait contains closed trajectories and particles on these trajectories are trapped around the resonance (i.e. cross $P=0$ many times). The growth of the area $S$ filled by such trapped trajectories in $(\zeta, P)$ plane implies that  particles from the region with open trajectories can be trapped into the resonance. The area $S$ reaches its maximum at a value of $s$ where $|u/r|$ is maximal;  as  $s$ further decreases, this area starts decreasing. Such area $S$ decrease corresponds to particles detrapping (escape) from the resonance. Close to the equator $s=0$ all particles will be detrapped and only open trajectories remain in the phase portrait (see Fig. \ref{fig3}).

The evolution of the area $S$ with closed trajectories along $s$ is much slower than trapped particles rotation around $P=0$, and thus this periodic rotation has an adiabatic invariant $I_\zeta=(2\pi)^{-1}\oint{Pd\zeta}$. At the moment of trapping $I_\zeta$ equals to $S/2\pi$ (and $S$ increases), and due to $I_\zeta$ conservation the detrapping should appear when $S$ decreases and comes back to the value $2\pi I_\zeta$. However, during this $S$ evolution the resonant energy would change (see the example of $\gamma_R$ profile along $s$ in Fig. \ref{fig3}). Thus, between trapping and detrapping electrons gain energy $\Delta\gamma_{trap}$, and this energy gain is determined by $S(\gamma_R)$ curve $ S = \oint {P_\zeta  d\zeta }  $:
\begin{equation}
S  = 2\sqrt {\frac{{2\varepsilon }}{w}} \int\limits_{\zeta _ -  }^{\zeta _ +  } {\sqrt {\left( {r\zeta _ +   + u\sin \zeta _ +  } \right) - \left( {r\zeta  + u\sin \zeta } \right)} d\zeta } \label{eq:10}
\end{equation}
where $\zeta_{\pm}$ are shown in Fig. \ref{fig3}. The amount of particles that will be trapped during a single resonant interaction (i.e., the probability of trapping, $\Pi$) is determined by the gradient of $S$ \citep[see][]{Neishtadt75, Neishtadt05, Shklyar81}: $\Pi = (\omega/2\pi m_ec^2)\cdot(\partial S/\partial \gamma)$.

All particles that are not trapped cross the resonance $P=0$ once, moving along open trajectories. These particles are scattered with the energy change $\Delta\gamma=\omega\Delta I/m_ec^2$, which depends on the phase $\zeta_R$ at the resonance:
\begin{eqnarray}
 \Delta \gamma  &=& -\int\limits_{ - \infty }^{\zeta _R } {\frac{{2\sqrt {\frac{{2\omega ^2 }}{{m_e c^2 g}}} \sqrt {\frac{{2I_R \Omega _0 }}{{m_e c^2 }}} \frac{{eB_w }}{{k\gamma _R m_e c^2 }}\cos \zeta }}{{\sqrt {H_\zeta   - \left\{ {\Lambda ,I_R } \right\}\zeta  - \sqrt {\frac{{2I_R \Omega _0 }}{{m_e c^2 }}} \frac{{eB_w }}{{k\gamma _R }}\sin \zeta } }}} d\zeta  \nonumber\\ 
  &=&  - 2\sqrt {\frac{{2\varepsilon \omega ^2 }}{{m_e^2 c^4 g}}} \int\limits_{ - \infty }^{\zeta _R } {\frac{{u\cos \zeta }}{{\sqrt {2\pi \theta r  - r\zeta  - u\sin \zeta } }}} d\zeta  \label{eq:11} 
 \end{eqnarray}
where $2\pi\theta=\zeta_R+(u/r)\sin\zeta_R\; ({\rm mod}\;2\pi)$. The function $\Delta\gamma(\theta)$ is periodic; its mean value over $\theta\in[0,1]$ is equal to $\Delta\gamma_{scat}=\langle\Delta\gamma(\theta)\rangle_\theta=-(\omega/2\pi m_ec^2)\cdot S$ \citep[see][]{Neishtadt99,Artemyev18:cnsns}. Thus, for the phase portrait with $S\ne 0$ the energy change due to scattering is also determined by the $S(\gamma_R)$ profile. This relation has been verified many times with numerical simulations for electron interactions with whistler-mode waves \citep[see][]{Albert13:AGU, Artemyev15:pop:probability, Vainchtein18:jgr}. For the phase portrait with $S=0$ we get $\langle\Delta\gamma(\theta)\rangle_\theta =0$ (i.e. no energy drift due to the resonance), but $\langle\left(\Delta\gamma(\theta)\right)^2\rangle_\theta \ne 0$ and there is an energy diffusion. Equation (\ref{eq:11}) shows that for $|u/r|\ll 1$ we get
\begin{equation}
\Delta \gamma  \approx  - 2\varepsilon \sqrt {\frac{{2\omega ^2 }}{{m_e c^2 g}}} \int\limits_{ - \infty }^{\zeta _R } {\frac{{u\cos \zeta d\zeta }}{{\sqrt {2\pi \theta  - \left\{ {\Lambda ,I_R } \right\}\zeta /m_e c^2 } }}} 
\label{eq:12}
\end{equation}
and $D_{QL}=\langle\left(\Delta\gamma(\theta)\right)^2\rangle_\theta \sim \varepsilon^2$, as it should be for quasi-linear diffusion coefficients proportional to the wave intensity (see detailed comparison of $D_{QL}$ calculated with Eq. (\ref{eq:12}) and quasi-linear diffusion rates in \citep{Karpman74:ssr, Albert10}).

\begin{figure}
\centering
\includegraphics[width=0.48\textwidth]{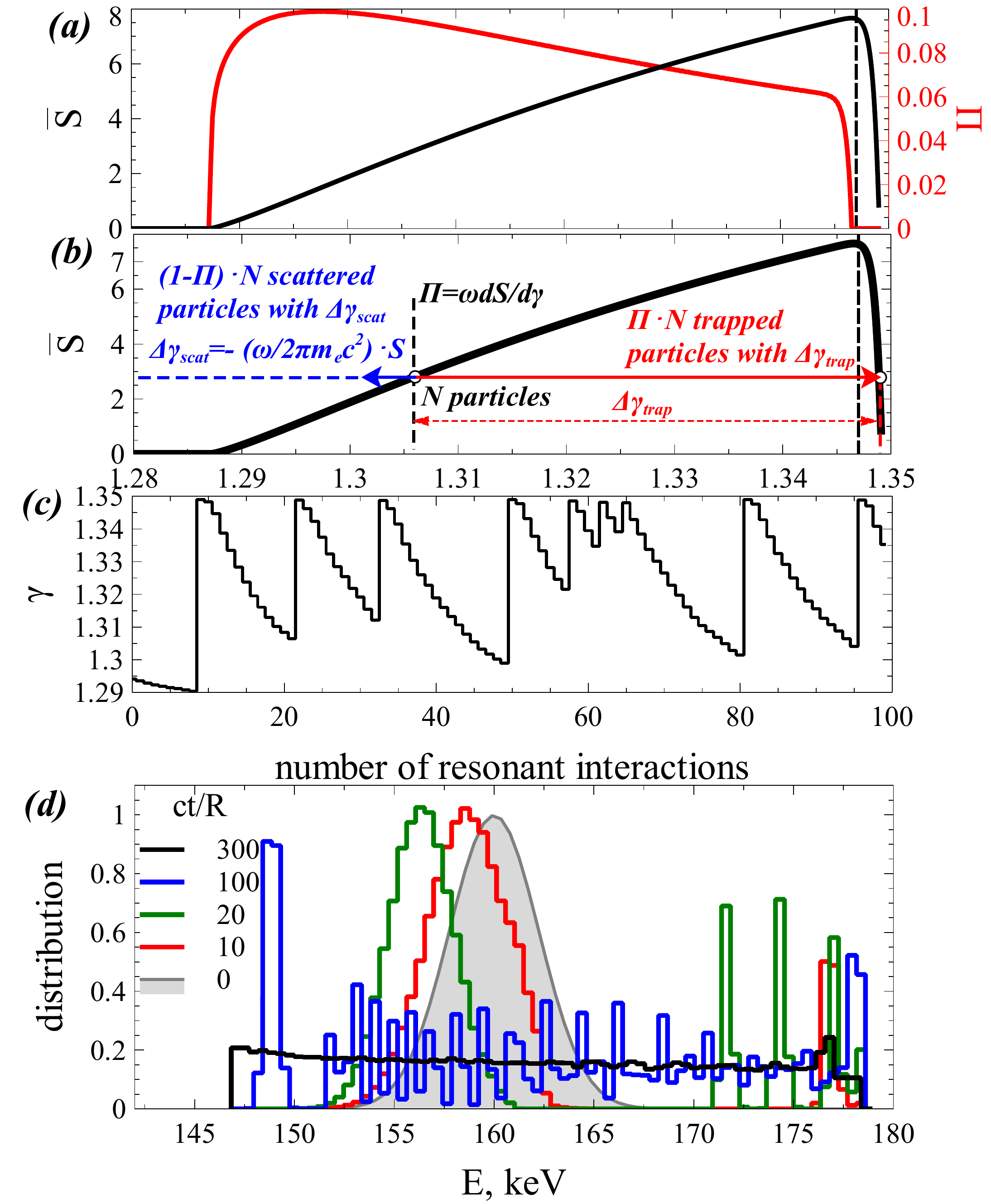}
\caption{
\label{fig4} Panel (a) shows profiles of $S(\gamma)$, $\Pi(\gamma)$ for the trajectory from Fig. \ref{fig1}(b). Schematic view of electron motion along $\gamma$ for given $\bar{S}(\gamma)=S\cdot(R\Omega_0(0)/c)^{1/2}$ (b). (note $R\Omega_0(0)/c$ is of the order of $1/\varepsilon$ for the system parameters). An example of trajectory $\gamma(n)$ for $n=0..100$ resonant iterations (c). The relaxation of the distribution function $f(\gamma)$ (d). Plots in panels (c,d) are obtained with map (\ref{eq:13}). }
\end{figure}

As electron energy changes only at resonances and remains constant between resonances, its evolution for any trajectory can be described by the mapping technique \citep[e.g.,][]{Chirikov78}. For the nonlinear resonances such a map takes the form \citep[see][]{Artemyev20:pop}:
\begin{equation}
\gamma _{n + 1}  = \gamma _n  + \left\{ {\begin{array}{*{20}c}
   {\Delta \gamma _{trap} ,} & {\theta  \in \left[ {0,\Pi } \right]}  \\
   {\Delta \gamma _{scat} ,} & {\theta  \in \left( {\Pi ,1} \right]}  \\
\end{array}} \right.
\label{eq:13}
\end{equation}
where $n$ is the number of resonance crossings (the number of map iterations). Equation (\ref{eq:13}) shows that new energy (after $n+1$ resonant interaction) $\gamma_{n+1}$ is the sum of energy $\gamma_n$ after previous resonance and energy change due to scattering or trapping. Variable $\theta$ is the rescaled effective energy at the resonance (see Eq. (\ref{eq:11})), and the variation of $\theta$  between two resonance crossings is determined by the normalized phase gain $\Delta\zeta\sim \varepsilon^{-1} \gg 1$ \citep[see equations in Appendix of][]{Artemyev20:pop}. This energy $\theta$ can be treated as a random variable with uniform distribution on $[0,1]$ \citep[see discussion of possible limitations of such a consideration in][]{Artemyev20:rcd}. Map (\ref{eq:13}) shows that for fixed $h$ the entire dynamics of electron energy due to nonlinear resonant interactions is described by a single curve $S(\gamma)$: the energy change due to scattering is $\Delta\gamma_{scat}=-(\omega/2\pi m_ec^2)\cdot S$, the probability of trapping is $\Pi=(\omega/2\pi m_ec^2)\cdot(dS/d\gamma)$, and energy change due to trapping $\Delta\gamma_{trap}$ is defined as the difference of $\gamma_R$ between two $S(\gamma)$ values (see scheme in Fig. \ref{fig4}(b)). 

To verify the mapping technique described by Eq. (\ref{eq:13}) we calculate $S(\gamma)$ for $h$ from Fig. \ref{fig1}(b) and iterate a large ensemble of trajectories. Figure \ref{fig4}(a) shows $S(\gamma)$, $\Pi(\gamma)$ profiles, several individual trajectories, and dynamics of $f(\gamma)$ relaxation. Note that to transform iterations $n$ into time, one should use $t_{n+1}=t_n+\tau_b(\gamma)$ where $\tau_b$ is the electron bounce period, i.e. the time interval between two resonant interactions (we consider waves only in the $s>0$ hemisphere): for the dipole field $\tau_b\approx 4(1-\gamma^{-2})^{-1/2}\left(1.3802-0.6397\sin^{3/4}\alpha_0\right)$ \cite{Orlova11}. Comparison of Figs. \ref{fig2} and \ref{fig4} demonstrates the applicability of the mapping technique for the description of electron resonant dynamics \citep[see more examples with this approach in ][]{Artemyev20:pop, Artemyev21:jpp}.

To summarize, the nonlinear resonances provide trapping and nonlinear scattering with energy changes described by $S(\gamma)$ curve. In this regime of resonant interaction $S\sim \sqrt{\varepsilon}$ (see Eq. (\ref{eq:10})). As $|u/r|$ decreases, the magnitude of $S$ tends to zero, and for sufficiently small $|u/r|$ there is only diffusion of electrons with the diffusion rate $D_{QL}=\langle(\Delta\gamma)^2\rangle\sim \varepsilon^2$ (see Eq. (\ref{eq:12}) and \citep{Albert10}). We aim to investigate the transition between these two regimes and to obtain estimates of the corresponding different time-scales of $f(\gamma)$ relaxation.

\section{Main system scalings}\label{sec:scalings}
To consider the transition between nonlinear resonant interaction and quasi-linear diffusion, we need to consider a $S\to 0$ limit for a fixed $\varepsilon$. We start with the evaluation of Hamiltonian (\ref{eq:09}) with small $S$. Next, we investigate time-scales of $f(\gamma)$ relaxation for $S\to 0$. Then we explain the relation of these time-scales and the time-scale of electron acceleration within the trapping. Finally, all obtained equations are combined together to provide the relation between time-scales of $f(\gamma)$ relaxation due to nonlinear resonances and quasi-linear diffusion.

\subsection{Threshold $S$ values}
\begin{figure} 
\centering
\includegraphics[width=0.48\textwidth]{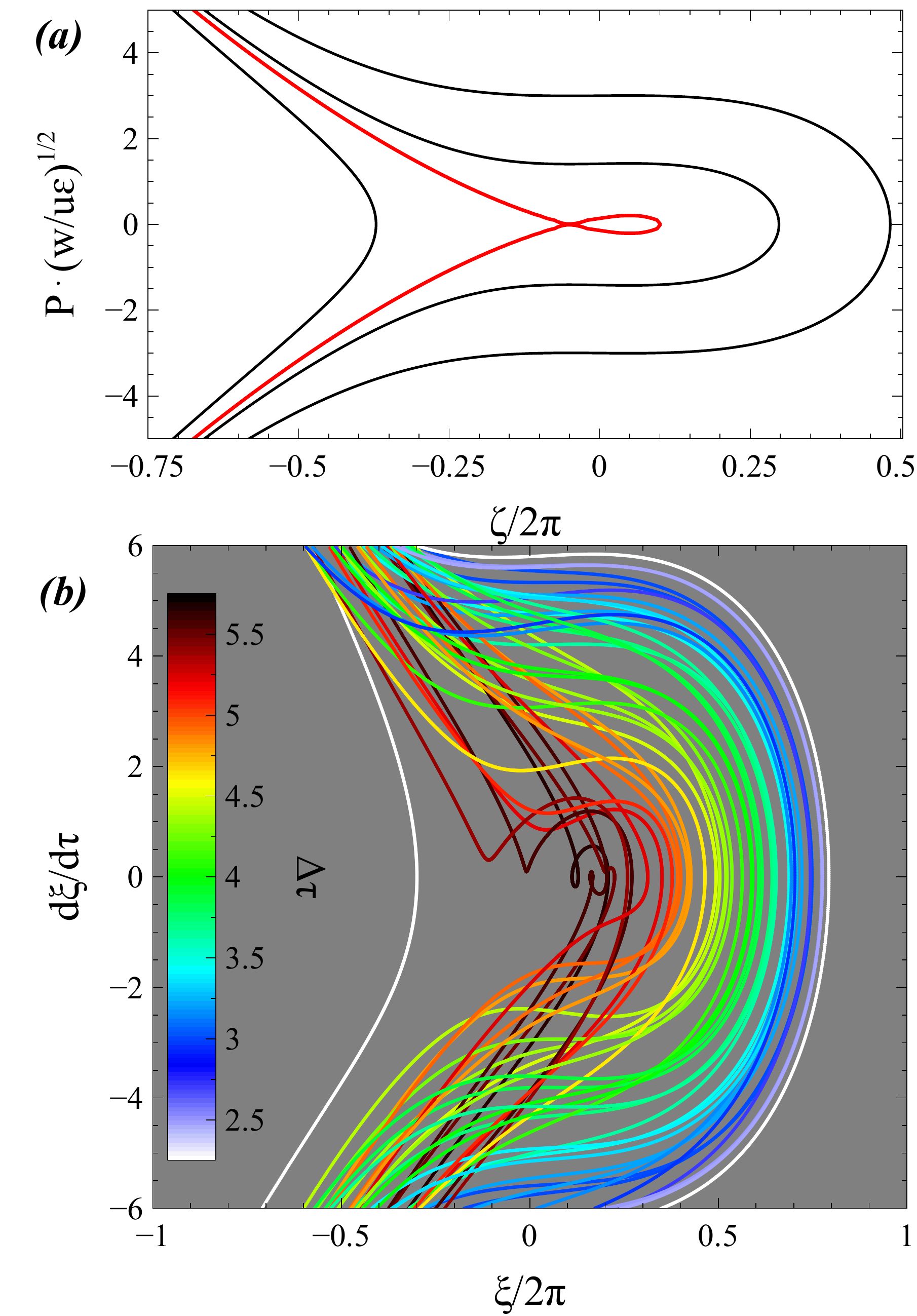}
\caption{
\label{fig5} The phase portrait of Hamiltonian (\ref{eq:14}) for $r/u$ around $1$ (a). Set of trajectories obtained by numerical integration of Eq. (\ref{eq:18}) with $\delta_0=\varepsilon^{2/3}$. Color codes the time interval required to cross the shown $d\xi/d\tau$ range.}
\end{figure}

Let us consider Hamiltonian (\ref{eq:09}) for $|r/u|$ around one, i.e. when $|r|$ becomes smaller than $u$ (condition required for existence of $S\ne 0$) only within a short $\gamma_R$ range. Then the $\zeta$ range of the area filled with closed trajectories ($\zeta \in[\zeta_-,\zeta_+]$ in Fig. \ref{fig3}(b)) is small and Hamiltonian (\ref{eq:09}) can be expanded as:
\begin{eqnarray}
 \Ham = \frac{1}{2}wP^2  + \varepsilon u \cdot \left( {\frac{r}{u}\zeta  + \sin \zeta } \right) \nonumber\\ 
\label{eq:14}\\
  \approx \frac{1}{2}wP^2  + \varepsilon u \cdot \left( {\left( {\frac{r}{u} - 1} \right)\zeta  + \frac{1}{6}\zeta ^3 } \right) \nonumber
 \end{eqnarray}
where $r/u$ slowly changes along the trajectory from $>1$ (no closed trajectories in the phase portrait) to $\min r/u<1$ (maximum area filled by closed trajectories), and then to $>1$ again. Figure \ref{fig5}(a) shows the phase portrait of Hamiltonian (\ref{eq:14}) for such $r/u<1$. To model this evolution we can write $r/u=1-\delta_0+(\varepsilon t)^2$ where the small parameter $\delta_0>0$ determines how far $|r/u|$ is from $1$ and $(\varepsilon t)^2$ models the slow evolution of $r/u$ along the trajectory (i.e. we change the slow coordinate $s$ to a slow time here):
\begin{equation}
\Ham \approx \frac{1}{2}wP^2  - \varepsilon u \cdot \left( {\left( {\delta _0  - \left( {\varepsilon t} \right)^2 } \right)\zeta  - \frac{1}{6}\zeta ^3 } \right)
\label{eq:15}
\end{equation}
and $w$, $u$ can be considered as constant along a short interval of $\varepsilon t \in [-\sqrt{\delta_0}, \sqrt{\delta_0}]$. The equation for $S$ for Hamiltonian (\ref{eq:15}) can be written as
\begin{eqnarray}
 S &=& \sqrt {\frac{{8u\varepsilon }}{w}} \int\limits_{\zeta _ +  }^{\zeta  - } {\left( {\left( {\delta _0  - \left( {\varepsilon t} \right)^2 } \right)\left( {\zeta   - \zeta _ + } \right) - \frac{{\zeta ^3  - \zeta _ + ^3 } }{6}} \right)^{1/2}d} \zeta \nonumber \\ 
  &=& \sqrt {\frac{{8u\varepsilon }}{w}} \delta _0^{5/4}\left( {1 - \frac{{t^2 }}{{t_0^2 }}} \right)^{5/4} \nonumber \\ 
  &\times& \int\limits_{\bar \zeta _ +  }^{\bar \zeta  - } {\left( {\left( {\bar \zeta   - \bar \zeta _ + } \right) - \frac{{\bar \zeta^3  - \bar \zeta _ +  ^3 }}{6}} \right)^{1/2}d} \bar \zeta  \nonumber \\ 
  &=& \sqrt {\frac{{8u\varepsilon }}{w}} \delta _0^{5/4} \left( {1 - \left( {t/t_0 } \right)^2 } \right)^{5/4}\cdot\frac{{12 \cdot 2^{3/4} }}{5}
 \label{eq:16}
 \end{eqnarray}
where $\bar\zeta_+=-\sqrt{2}$, $\bar\zeta_-=\sqrt{8}$, $t_0=\sqrt{\delta_0}/\varepsilon$ and $t/t_0$ is equivalent to $\gamma_R$, the energy at resonance, because different slow time values imply here different values of slow $s$, i.e. different values of resonant $s_R$ related to $\gamma_R$ through Eq. (\ref{eq:05}) and $h={\rm const}$. Equation (\ref{eq:16}) demonstrates that the $\delta_0$ parameter controls the magnitude of $S$, and effects of nonlinear interactions should disappear as $\delta_0\to 0$. In the nonlinear regime, there are well separated populations of trapped particles (a small number of particles gaining a large energy) and nonlinearly scattered particles (a large number of particles losing energy). In contrast, in the diffusive regime the numbers of particles gaining and losing energy are (approximately) equal. Therefore, there is a threshold $\delta_0$ value (or $S$ value) such that for $\delta_0$ below this threshold, we cannot separate trapping and nonlinear scattering. Let us obtain this threshold $\delta_0$ (or $S$) value. 

Hamiltonian equations for Hamiltonian (\ref{eq:15}) can be combined to get the second order equation for $\zeta$:
\begin{equation}
\frac{{d^2 \zeta }}{{d\tilde t^2 }} =   \varepsilon  \cdot \left( {\delta _0  - \left( {\varepsilon \tilde t} \right)^2  - \frac{1}{2}\zeta ^2 } \right)
\label{eq:17}
\end{equation}
where $\tilde{t}=t\sqrt{uw}$ and $u$, $w$ are constants of the order of one (not dependent on $\varepsilon$). Equation (\ref{eq:17}) can be rewritten in normalized variables $\tau=\tilde{t}\varepsilon^{1/2}\delta_0^{1/4}$, $\xi=\zeta/\sqrt{\delta_0}$
\begin{equation}
\frac{{d^2 \xi }}{{d\tau ^2 }} =   \left( {1 - \frac{\varepsilon }{{\delta _0^{3/2} }}\tau ^2 } \right) - \frac{1}{2}\xi ^2 
\label{eq:18}
\end{equation}
Equation (\ref{eq:18}) shows that for $\delta_0=\varepsilon^{2/3}$ there is no separation of time-scales, i.e. the equation describing the electron motion around the resonance does not contain a slow time. Thus, both trapped and nonlinearly scattered particles should stay in the resonance approximately the same time, and there is no separation between these two types of trajectories. Figure \ref{fig5}(b) confirms this conclusion: we solve equation Eq. (\ref{eq:18}) with $\delta_0=\varepsilon^{2/3}$ for set of trajectories and plot these trajectories in the phase plane $(\xi,d\xi/d\tau)$. There are still some trajectories similar to trapped trajectories of the original system, but particles on these trajectories do not fulfil a single oscillation across the resonance $d\xi/d\tau=0$, i.e. we cannot separate particles on trapped and scattered trajectories. 

\subsection{Time-scales of $f(\gamma)$ evolution}
\begin{figure*}
\centering
\includegraphics[width=0.95\textwidth]{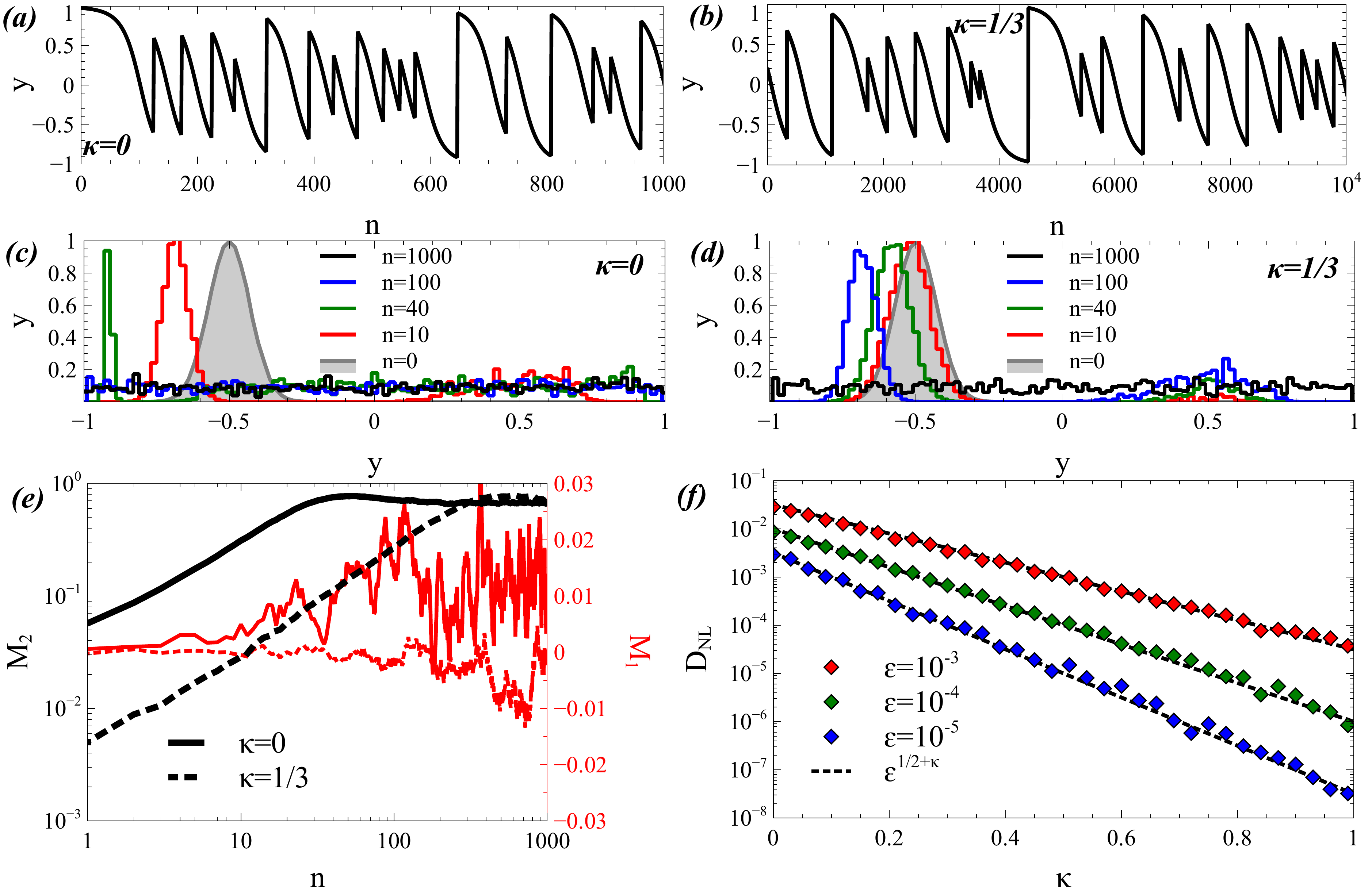}
\caption{
\label{fig6} Examples of $y_n$ trajectories described by map (\ref{eq:19}) (a,b) and evolution of $f(y)$ distribution (c,d) for two $\kappa$ values. Profiles of $\M_1(n)$, $\M_2(n)$ for two $\kappa$ values (e). Dependence of $D_{NL}$ on $\kappa$ for three $\varepsilon$ values (f).  }
\end{figure*}

To investigate how the rate of $f(\gamma)$ evolution scales with $\delta_0$ we use the mapping technique. Equation (\ref{eq:16}) shows that the area $S$ can be set as $\varepsilon^{1/2}\delta_0^{5/4}\cdot(1-y^2)^{5/4}$ where $y$ is the effective particle energy at the resonance (shifted relative to the energy where $S$ reaches maximum), i.e. different values of slow time $\sim \varepsilon t$ in Eq. (\ref{eq:16}) correspond to  different $s_R$ values of the original system and to different resonant energies. Therefore, we can rewrite mapping (\ref{eq:13}) as
\begin{equation}
y_{n + 1}  = y_n  + \left\{ {\begin{array}{*{20}c}
   { - 2y_n ,} & {\theta  \in \left[ {0,\Pi } \right]}  \\
   { - \s\left( {y_n } \right),} & {\theta  \in \left( {\Pi ,1} \right]}  \\
\end{array}} \right. \label{eq:19}
\end{equation}
and
\begin{eqnarray}
 \s &=& \varepsilon ^{1/2 + \kappa } \left( {1 - y^2 } \right)^{5/4} \nonumber \\ 
 \Pi  &=& \frac{{d\s}}{{dy}} =  - \frac{5}{2}y\varepsilon ^{1/2 + \kappa } \left( {1 - y^2 } \right)^{1/4}  \nonumber
\end{eqnarray}
where we introduce $\kappa$ as $\delta_0=\varepsilon^{4\kappa/5}$, and take into account $\s(-y)=\s(y)$ to calculate $y$ change due to trapping. Figure \ref{fig6}(a) shows an example of $y_n$ trajectory described by map (\ref{eq:19}). This trajectory repeats all features seen in the original system trajectories (see Figs. \ref{fig1}(d), \ref{fig4}(c)). The relaxation of the initially localized $f(y)$ distribution also repeats well the relaxation of the $f(\gamma)$ distribution: compare Fig. \ref{fig6}(b) and Figs. \ref{fig2}(d), \ref{fig4}(d).

To characterize the time-scale of $f(y)$ evolution as a function of $\kappa$, we numerically integrate a large ensemble of trajectories described by map (\ref{eq:19}) for various values of $\kappa$. The initial $y_{0,i}$ values for $i=0…N$ trajectories are uniformly distributed within $[-1,1]$ range, and we calculate two characteristics: $\M_1(n)=N^{-1}\sum_{i=0..N}\left(y_{n,i}-y_{0,i}\right)$ and $\M_2(n)= N^{-1}\sum_{i=0..N}\left(y_{n,i}-y_{0,i}\right)^2-\M_1^2(n)$. Figure \ref{fig6}(e) shows $\M_1(n)$, $\M_2(n)$ profiles for two $\kappa$ values: $\kappa=0$ corresponds to the system with well distinguished populations of trapped and scattered particles (i.e., the system is far from the diffusive regime), whereas $\kappa=1/3$ corresponds to $\delta_0=\varepsilon^{4\kappa/5}=\varepsilon^{4/15}$ and this is a case with much closer time-scales of the dynamics of trapped and scattered particle populations. For both values of $\kappa$  $\M_2$ grows with $n$ and $\M_1$ oscillates around the zero.
%For $\kappa=0$ we have quite significant $\M_1$, whereas $\M_2$ dominates for $\kappa=5/6$. \todo{text}

To derive the time-scale of $f(y)$ relaxation we fit the growing fragment of $\M_2(n)$ profile by the linear function $D_{NL}\cdot{n}$ with the coefficient $D_{NL}$ playing the role of a diffusion coefficient in systems with diffusive resonant interactions. Figure \ref{fig6}(f) shows the $D_{NL}$ dependence on $\kappa$ for three $\varepsilon$ values. There is a clear scaling $D_{NL}\approx \varepsilon^{1/2+\kappa}$, i.e. $D_{NL}$ varies as $\varepsilon^{1/2}$ for $\kappa=0$ and as $\varepsilon^{4/3}$ for $\kappa=5/6$. This scaling may be explained by the dominating role of trappings in the $\M_2$ increase. Most of nonlinearly scattered particles change their energy only slightly, and the first order contribution of this energy change goes into $\M_1$, whereas the contribution of these particles to the growth of $\M_2$is about $\sim \s^2 \sim \varepsilon^{1+2\kappa}$. However, even a small population of trapped particles would contribute to $ N^{-1}\sum_{i=0..N}\left(y_{n,i}-y_{0,i}\right)^2$ as $\sim \Pi\sim \varepsilon^{1/2+\kappa} $ (the number of trapped particles is about the probability of trapping, whereas these particles have $y_{n,i}-y_{0,i}\sim O(1)$); on the other hand, their contribution to $\M_1^2$ is much smaller. Thus, $\M_2\sim \Pi$ and this confirms the scaling from Fig. \ref{fig6}(f).

\subsection{Time-scale of trapping acceleration}
The scaling from Fig. \ref{fig6}(d) shows that $D_{NL}$ not only depends on $\varepsilon$, but also scales with $\delta_0\sim \varepsilon^{4\kappa/5}$. This parameter, $\delta_0$, controls the effective range of energy change due to trapping, and should be related to the number of complete rotations around the resonance of trapped particles, $N_{trap}$. Thus, we can derive the $\delta_0(N_{trap})$ dependence and rewrite $D_{NL}$ as a function of $\varepsilon$ and $N_{trap}$. We define $N_{trap}$ as
the maximum number of periods of a trapped particle's rotation in the phase space (see phase portrait in Fig. \ref{fig3}(b)). This parameter is quite universal: it can be determined for any particular wave-field model (see, e.g., \cite{Zhang18:jgr:intensewaves} for a discussion of $N_{trap}$ values typical for whistler-mode waves observed in the Earth's radiation belts). The trapped period for Hamiltonian (\ref{eq:15}) can be written as
\begin{equation}
T_{trap} \left( {\varepsilon t} \right) = 2\int\limits_{\zeta _ -  }^{\zeta _{_ +  } } {\frac{{d\zeta }}{{\dot \zeta }}}  = \frac{C}{{\varepsilon ^{1/2} \left( {\delta _0  - \varepsilon ^2 t^2 } \right)^{1/4} }}
\label{eq:20}
\end{equation}
where constant $C\sim O(1)$ is determined by the distance from the separatrix in the phase portrait. The maximum time of trapped particle motion is of the order of $\sim2\delta_0^{1/2}/\varepsilon$. Therefore we can write for $N_{trap}$
\begin{equation}
N_{trap}  = \frac{2}{\varepsilon }\int\limits_0^{\sqrt {\delta _0 } /\varepsilon } {\frac{{d\varepsilon t}}{{T_{trap} \left( {\varepsilon t} \right)}}}  = \frac{{\delta _0^{3/4} }}{{\sqrt \varepsilon  }}\tilde C\sim\varepsilon ^{3\kappa /5 - 1/2} \label{eq:21}
\end{equation}
where constant $\tilde{C}\sim O(1)$. Equation (\ref{eq:21}) shows that $N_{trap}\sim \varepsilon^{3\kappa/5-1/2}$. For the threshold value $\kappa=5/6$, we get $N_{trap}\sim O(1)$, i.e. indeed for $\kappa=5/6$ ($\delta_0\sim \varepsilon^{2/3}$) the number of trapping periods does not depend on $\varepsilon$ and there is no separation between trapped and scattered particles anymore. 

Combining Eq. (\ref{eq:21}) and the $D_{NL}$ scaling, we obtain the final scaling of the coefficient $D_{NL}$
\begin{equation}
D_{NL} \sim\varepsilon ^{1/2 + \kappa } \sim \varepsilon ^{4/3} N_{trap}^{5/3} \sim D_{QL}  \cdot \varepsilon ^{ - 2/3} N_{trap}^{5/3} 
\label{eq:22}
\end{equation}
where we take into account that $D_{QL}\sim \varepsilon^{2}$. Therefore, $D_{NL}/D_{QL}=\Q\varepsilon^{-2/3}N_{trap}^{5/3}$ where the factor $\Q\sim O(1)$ is determined by the system parameters. According to this scaling, the transition between the diffusive regime ($N_{trap}\to 0$) and the nonlinear resonant regime ($N_{trap}\gg 1$) occurs when $N_{trap}\sim \varepsilon^{2/5} \ll 1$, whereas for $\kappa=5/6$ ($N_{trap}\sim 1$) we get $D_{NL}/D_{QL}\sim \varepsilon^{-2/3}\gg1$.

However, the scaling in Eq. (\ref{eq:22}) is obtained for systems where $\varepsilon$ and $N_{trap}$ (or more precisely $\kappa$, which determines $\delta_0$) are assumed to be independent parameters that can be varied separately. But $\varepsilon$ and $N_{trap}$ are not always independent. This is especially the case in realistic systems, where whistler-mode waves often propagate in the form of short wave packets (see examples of such wave packets in observations \cite{Zhang18:jgr:intensewaves} and in numerical simulations \cite{Nunn21, Zhang21:jgr:data&model}). We consider this realistic situation in the next subsection. 

\subsection{Systems with short wave packets: Shrinking of the resonant energy range}
In the Earth's magnetosphere, intense whistler-mode waves contain a fine structure of packets or subpackets, where packets/subpackets are formed by strong wave amplitude modulations (by at least a factor of $\sim 2$) accompanied by large and random wave phase jumps in-between successive packets/subpackets, allowing to treat electron interactions with different packets/subpackets independently \citep{Tao12:GRL, Tao13, Mourenas18:jgr, Zhang20:grl:phase}. In particular, most long wave envelopes, characterized by an amplitude continuously remaining above $50$ pT, actually consist of many shorter packets/subpackets when internal strong amplitude modulations and wave frequency/phase jumps are taken into account \citep{Mourenas18:jgr, Zhang20:grl:frequency, Zhang20:grl:phase}. The overwhelming majority of such intense parallel propagating whistler-mode wave packets/subpackets observed in the magnetosphere have a short length of $\beta<10$ wave periods, with much more rare long packets/subpackets reaching $\beta\sim 100-300$ periods \citep{Zhang18:jgr:intensewaves,Zhang19:grl,Mourenas18:jgr}. Short packets of length $\beta<10$ are likely often produced by wave superposition \citep{Zhang19:grl,Zhang20:grl:frequency,Nunn21} and their length $\beta$ is mostly independent of their amplitude $\varepsilon$. Packets of intermediate length $\beta\sim10-30$ are probably partly formed by trapping-induced amplitude modulation that depends on $\varepsilon^{-1/2}$, but statistical observations show that all packets are distributed over a wide range of $\beta$ values for any given $\varepsilon$ \citep{Tao13,Zhang20:grl:frequency, Nunn21}, allowing to treat $\beta$ and $\varepsilon$ as independent parameters to leading order. 

Taking into account that first-order cyclotron resonance between whistler-mode waves and $<1$ MeV electrons occurs for waves and electrons moving in opposite directions, the time interval of trapped particle motion is limited to $\sim 2\pi\beta/\Omega_{0}$, whereas one period of trapped electron oscillation is $T_{trap}\sim 2\pi\varepsilon^{-1/2}/\Omega_{0}$ for typical resonant energies \cite{Karpman74:ssr}. Therefore, $N_{trap}$ can be expressed as a function of the independent parameters $\beta$ and $\varepsilon$ of short wave packets, giving $N_{trap}\sim \beta\varepsilon^{1/2}$ \citep[see also][]{Zhang18:jgr:intensewaves}.

In the case of short wave packets, although the magnitude of $S$ remains $\sim \varepsilon^{1/2}$, the limited packet size can lead to important changes as compared with the scaling obtained in Eq. (\ref{eq:22}) of the preceding subsection in the ideal case of {\it infinitely long} wave packets or waves. Indeed, a given short wave packet occupies only a limited latitudinal extent along a magnetic field line. Eq. (\ref{eq:05}) and the conservation of $h$, which relate latitude of resonance to electron energy, imply that this limited latitudinal extent directly corresponds to a limited energy range for electron resonance with this packet during one bounce period. This limited energy range of actually resonant electrons represents a fraction $\ell<1$ of the total energy range of particles potentially reaching resonance with a wave packet over the full length of the magnetic field line. Electron trapping by a short packet leads to a smaller energy change than trapping by an ideal {\it infinitely long} wave packet, because the electron is released from trapping faster, corresponding to a reduction of its energy change by a factor $\sim\ell$. Besides, short wave packets may sometimes be rare and occur only once every bounce period, corresponding to a reduction of their occurrence rate by a factor $\sim\ell<1$ as compared with both the case of a close succession of short packets and the ideal case of {\it infinitely long} packets assumed in the preceding subsection.

\begin{figure*}
\centering
\includegraphics[width=0.95\textwidth]{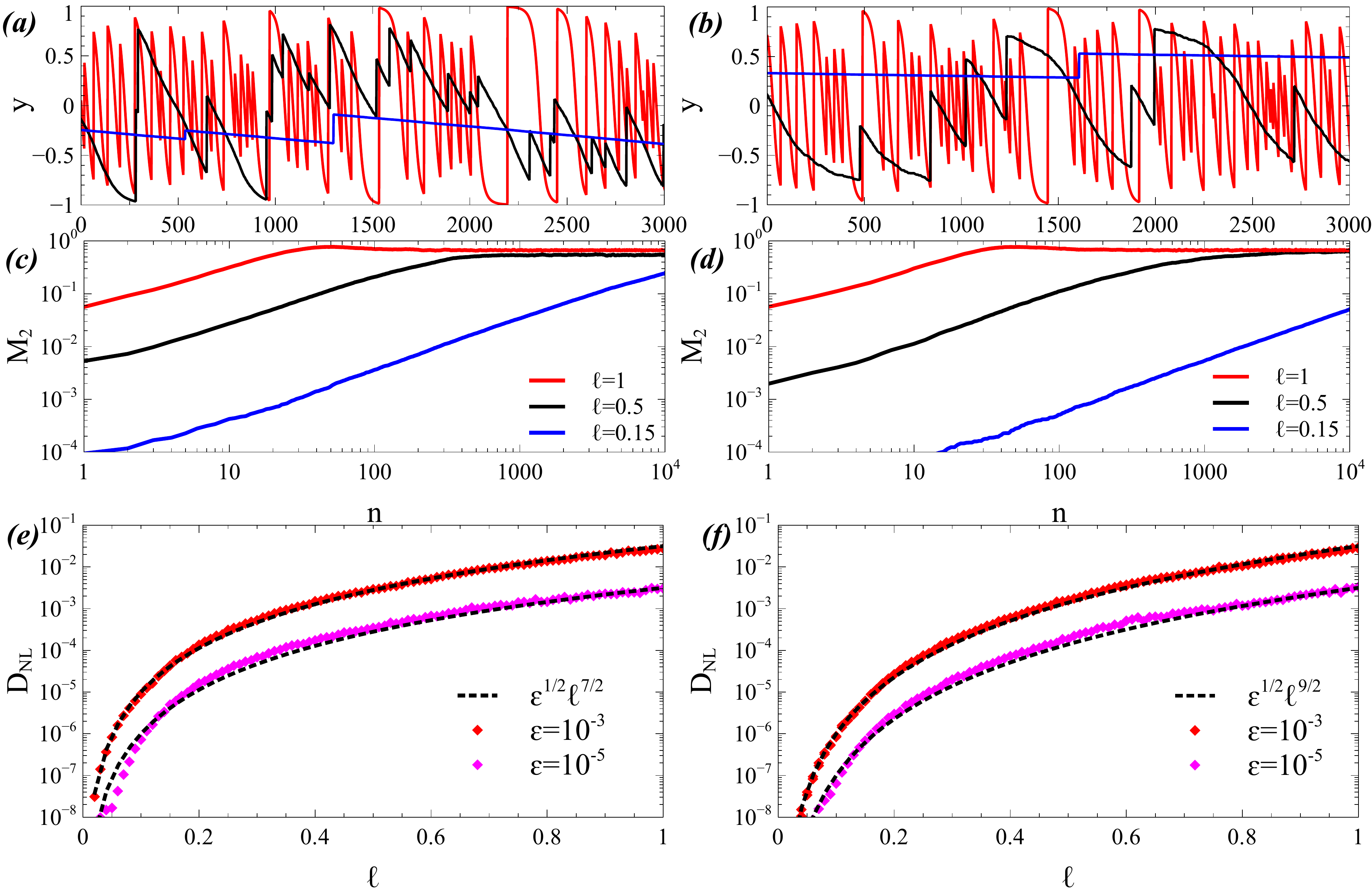}
\caption{
\label{fig7} Three trajectories for different $\ell$ values (see color codes in panels (c,d)) for systems of two different kinds (a,b). Examples of $M_{2}$ profiles for these two systems (c,d). Dependencies of $D_{NL}$ on $\ell$ for the two systems and different $\varepsilon$ (d,f). Left column (a,c,e) shows results for the system of the first kind: the entire resonant range $y\in[-1,1]$, but $S$ is not equal to zero only for $[-\ell, \ell]$ appearing always around the particle location, i.e. there is a resonant interaction at each iteration. The right column (b,d,f) shows results for the system of the second kind: the entire resonant range $y\in[-1,1]$, but $S$ is not equal to zero only over $[-\ell, \ell]$ appearing at random locations, not necessarily near the particle location, i.e., there are  iterations without change of particle $y$. }
\end{figure*}

Accordingly, let us consider a realistic situation where the magnitude of $\s\sim \sqrt{\varepsilon}$ remains the same, but the range of nonlinear resonant interactions (formally the $\varepsilon t$ range in Eq. (\ref{eq:15})) shrinks. There are two kinds of systems with an entire resonant range $y\in[-1,1]$, but where $\s$ is not equal to zero only for $y\in[-\ell, \ell]$, with $\ell=\sqrt{\delta_0}$. In systems of the first kind, we center $\s$ around a $\bar{y}_n$ randomly generated at each map iteration around $y_n$, i.e., for each iteration there is a finite change of $y_n$. This type of mapping mimics electron resonant interactions with a set of short wave packets that would fill the entire magnetic field lines, but which cannot trap electrons for a long time because of the limited packet duration. In the system of the second kind, we center $\s$ around a randomly generated $\bar{y}_n$, but we do not control the position of $\bar{y}_n$ relative to $y_n$ and some map iterations can occur without any change of $y_n$ because $y_n$ is outside the $\s\ne0$ range. This type of mapping mimics electron resonant interaction with rare short wave-packets propagating with some time separation. It corresponds to a situation where, during each bounce period, only one wave packet is present and many electrons reach the latitude of cyclotron resonance without encountering this intense packet there. For both systems, the mapping (\ref{eq:19}) can be rewritten as
\begin{equation}
y_{n + 1}  = y_n  + \left\{ {\begin{array}{*{20}c}
   {\bar y_n  - 2y_n ,} & {\theta  \in \left[ {0,\Pi } \right]}  \\
   { - \s_n \left( {y_n } \right) ,} & {\theta  \in \left( {\Pi ,1} \right]}  \\
\end{array}} \right.
\label{eq:23}
\end{equation}
and
\begin{eqnarray}
 \s_n  &=& \varepsilon ^{1/2} \left( {\ell ^2  - \left( {y - \bar y_n } \right)^2 } \right)^{5/4} \nonumber \\ 
 \Pi  &=& \frac{{d\s_n}}{{dy}} =  - \frac{5}{2}\left( {y - \bar y_n } \right)\varepsilon ^{1/2} \left( {\ell ^2  - \left( {y - \bar y_n } \right)^2 } \right)^{1/4}  \nonumber 
 \end{eqnarray}
where $\bar{y}_n=y_n+\ell\cdot R$ for the system of the first kind, and $\bar{y}_n=-1+\ell+2(1-\ell)\cdot R$ for the system of the second kind ; here $R$ is a random number with a uniform distribution within $[-1,1]$ (note for $\bar{y}_n=y_n+\ell\cdot R$ we also control that $\s\ne0$ range does not cross $y\pm1$ boundaries). An important property of both systems is that for small $\ell$ trappings are possible for any $y_n$ values, whereas for $\ell=1$ (the initial map given by Eq. (\ref{eq:19})) trapping is possible only for $y_n<0$. Figure \ref{fig7}(a,b) shows a set of sample trajectories for each of the two kinds of systems. The rate of change of $y$ is going down as $\ell$ decreases, and this effect is stronger for the system of the second kind  (Fig. \ref{fig7}(b)).

For these two kinds of systems we set the range of $\ell$ and for each $\ell$ value calculate $M_{2}(n)$. Figure \ref{fig7}(c,d) shows examples of $M_{2}$ profiles. We fit the growing fragment of $\M_2(n)$ by $D_{NL}n$, and  Fig. \ref{fig7}(e,f) shows the $D_{NL}$ dependence on $\ell$. For both systems, $D_{NL}$ scales with $\ell$ as $D_{NL}\sim\varepsilon^{1/2}\ell^\eta$. But $\eta\approx 7/2$ for the system of the first kind where resonances occur at each iteration, whereas $\eta\approx 9/2$ for the system of the second kind, where resonance occurrence is decreased by a factor $\ell\ll1$. Taking into account that $\ell=\sqrt{\delta_0}$, we can rewrite Eq. (\ref{eq:21}) as $N_{trap}\sim \ell^{3/2}/\sqrt{\varepsilon}$. Therefore, for $D_{NL}$ scaling with $N_{trap}$ we have $D_{NL}\sim \varepsilon^{1/2}\ell^\eta\sim \varepsilon^{1/2+\eta/3}N_{trap}^{2\eta/3}$ and
\begin{equation}
D_{NL} \sim\left\{ {\begin{array}{*{20}c}
   {\varepsilon ^{5/3} N_{trap}^{7/3} ,} & {\eta  = 7/2}  \\
   {\varepsilon ^2 N_{trap}^3 ,} & {\eta  = 9/2}  \\
\end{array}} \right.
\label{eq:24}
\end{equation}
for the two kinds of systems. 

A general scaling, including both the results in Eq. (\ref{eq:24}) and in Eq. (\ref{eq:22}), is 
\begin{equation}
D_{NL}\sim
   \varepsilon ^{1/2+\eta/3} N_{trap}^{2\eta/3}\sim \varepsilon ^{1/2+2\eta/3} \beta^{2\eta/3}.
\label{eq:25}
\end{equation}
For $\eta=5/2$, Eq. (\ref{eq:25}) allows to recover the initial scaling from Eq. (\ref{eq:22}), which corresponds to the ideal case of electron interactions with {\it infinitely long} wave packets/waves present all the time along magnetic field lines. In the more realistic situation of short wave packets, two different kinds of systems are possible, corresponding to $\eta=7/2$ and $\eta=9/2$. For the system of the first kind, we get $D_{NL}/D_{QL}\sim \varepsilon^{-1/3}N_{trap}^{7/3} \sim \varepsilon^{5/6}\beta^{7/3}$. In this case, $D_{NL}/D_{QL}\sim O(1)$ requires that $N_{trap}\sim \varepsilon^{1/7}$ and $\beta\sim \varepsilon^{-5/14}$. For realistic whistler-mode wave amplitudes $\varepsilon\in[3\times 10^{-4},3\times 10^{-3}]$ in the Earth's radiation belts \citep{Zhang20:grl:frequency}, the transition between the regimes of quasi-linear diffusion and nonlinear resonant interaction should therefore occur approximately for $N_{trap}>0.35$ and $\beta>10$. In the system of the second kind, we have $D_{NL}/D_{QL}\sim N_{trap}^{3}\sim \beta^3\varepsilon^{3/2}$. In this case, $D_{NL}/D_{QL}\sim O(1)$ corresponds to $N_{trap}\sim 1$ and $\beta\sim\varepsilon^{-1/2}$. Thus, for this system of very rare short packets, the transition between the regimes of quasi-linear diffusion and nonlinear resonant interaction should occur around $N_{trap}\sim 1$ and approximately for $\beta>20$ -- i.e. near the $\kappa=5/6$ threshold value when $\delta_0\sim\varepsilon^{2/3}$ and $\ell\sim \varepsilon^{1/3}$.

Note that the $D_{NL}$ scaling in Eq. (\ref{eq:25}) for the first kind of system with short wave packets, corresponding to $\eta=7/2$, is valid only when cyclotron resonance with wave packets is always available, i.e. when electrons with any energy within the resonant range interact with the wave once (or more frequently) per bounce period. In practice, however, intense wave packets are not always present \citep{Zhang18:jgr:intensewaves, Mourenas18:jgr}. Nevertheless, the overwhelming majority of wave packets interact with electrons independently of the other packets \citep{Zhang20:grl:phase}, and their very large population is mostly randomly distributed in time. Consequently, over a sufficiently long time period, all latitudes of resonance with all electron energies should be uniformly reached by packets in each ($\beta,\varepsilon$) range. In this limit, the time-averaged $\langle D_{NL}\rangle$ can be simply obtained by a double integration over the full measured distribution of $(\beta>0,\varepsilon>0)$ packets of $D_{NL}$, weighted by the measured temporal occurrence rate (taking into account packet duration $\beta$) of $(\beta,\varepsilon)$ packets \citep{Zhang18:jgr:intensewaves, Zhang19:grl, Mourenas18:jgr}. Finally, $\langle D_{NL}\rangle/\langle D_{NL}\rangle$ can be obtained by dividing $\langle D_{NL}\rangle$ by the integral over $\varepsilon$ of $D_{QL}\sim\varepsilon^2$ weighted by the measured temporal occurrence rate of $\varepsilon$ packets. In addition, $D_{NL}/D_{QL}$ should be considered as $\ne 1$ in the above integration only for $\varepsilon$ values above the threshold for nonlinear effects \citep{Omura08,Zhang19:grl}. For lower amplitudes, one can simply assume $D_{NL}/D_{QL}\approx 1$.

\section{Discussion and Conclusion}\label{sec:discusison}
In this study, we investigate the transition between two regimes of resonant electron interactions with whistler-mode waves: quasi-linear diffusion and nonlinear resonant interaction. The typical effects of these two regimes are very different: diffusive scattering results in small energy changes with zero mean value, whereas nonlinear resonant interactions result in energy drifts (due to phase bunching) or large jumps (due to phase trapping). This difference makes it impossible to directly compare energy changes for these two regimes, i.e. energy drifts and jumps cannot be compared with the diffusion rates. Therefore, a more general (nonlocal, i.e. independent on the initial energy value) characteristic should be determined to compare these two regimes. For such characteristic, we choose the typical time-scale of relaxation of the energy distribution function $f(\gamma)$, which is determined along a single resonant curve $h={\rm const}$. This characteristic describes the entire energy, pitch-angle range for a given $h=\gamma-\left(\omega/2\Omega(0)\right)(\gamma^2-1)\sin^2\alpha_0$ and wave frequency, i.e. it is an integral characteristic that simplifies the system description. The relaxation of the $f(\gamma)$ distribution depends on the initial $f(\gamma)$ and contains many elements that are different for these two regimes (e.g., formation of local maxima due to the phase trapping acceleration). Therefore, to further simplify the description of this evolution and reduce it to a single parameter, we restrict our consideration to the dynamics of $f(\gamma)$ dispersion $\M_2$ that increases with the number of resonance crossings and finally saturates. This parameter $\M_2(n)$ is the unique characteristic for the diffusive resonant regime, but for nonlinear resonant interactions there are also $\M_1(n)$ and higher-order moments. Thus, some justification for taking only $\M_2$ into account should be given. The first moment $\M_1(n)=Vn$ scales with $\varepsilon$ as $V\sim \s\sim \varepsilon^{1/2+\kappa}$, and the same scaling can be obtained for $\M_2(n)=D_{NL}n$ with $D_{NL}\sim \s\sim \varepsilon^{1/2+\kappa}$. The beginning of the evolution (relaxation) of the distribution $f(\gamma)$ is dominated by trapping effects contributing to $\M_2$, because $\delta\gamma^2\sim D_{NL}n$ is larger than $\delta\gamma^2\sim V^2n^2$ for small $n$. The threshold value of $n$  such that drifts become more important than trappings is $n^*\sim D_{NL}/V^2\sim \varepsilon^{-1/2-\kappa}$. During this number of resonant interactions the first trapped particles would be drifted back to their initial energies, because $\Delta\gamma_{trap}/\Delta\gamma_{scat}\sim 1/S\sim \varepsilon^{-1/2-\kappa}$ and $n^*\cdot\left(\Delta\gamma_{trap}/\Delta\gamma_{scat}\right)\sim O(1)$. The drift effects may control the relaxation of $f(\gamma)$ only after the main relaxation stage is finished (when the initially trapped particles reach back their initial energies, the first large cycle of trapping-scattering would be finished; see, e.g., example of such cycles in Figs. \ref{fig1}(d)). Thus, we indeed can use $\M_2(n)$ to characterize the main stage of $f(\gamma)$ relaxation. Note that for the regime of nonlinear resonant interaction $\M_2(n)$ does not describe diffusion, but shows how quickly the dispersion of $f(\gamma)$  grows due to trapping effects.  

In the Earth's magnetosphere, whistler-mode waves mainly propagate in the form of wave packets, and most of the observed parallel propagating intense wave packets are short, containing $\beta<10-30$ wave periods \citep{Zhang18:jgr:intensewaves, Mourenas18:jgr}. The relations obtained here, connecting $D_{NL}/D_{QL}$ with wave packet amplitude $\varepsilon$ and number of trapped electron oscillations $N_{trap}$ or packet length $\beta$, can be considered as inverse scaling laws for the time-scale of $f(\gamma)$ relaxation -- although such a consideration simplifies the details of $f(\gamma)$ evolution. The main potential application of these scaling laws is to allow including the leading order effect of nonlinear resonant interactions into global models of radiation belt dynamics relying on the Fokker-Planck diffusion equation. This equation describes the evolution of $f(\gamma)$ driven by diffusion on a typical time-scale $\sim 1/D_{QL}$. This time-scale is close to the observed time-scales of electron flux dynamics in the Earth's radiation belts during quiet time periods characterized by not-to-high wave intensity \citep[e.g.,][]{Artemyev13:angeo, Mourenas17, Claudepierre20:II}. However, during more disturbed periods with a high occurrence rate of intense wave packets \cite{Thorne13:nature}, the observed electron flux evolution could be faster than the evolution predicted by a diffusion model (see discussions in, e.g., \citep{Mourenas18:jgr, Gan20:grl:II, Allanson21}). Our scaling laws of $D_{NL}/D_{QL}$ allow for a simple renormalization of the quasi-linear diffusion rates based on the observed occurrence rate of intense waves resonantly interacting with electrons in the nonlinear regime. An important ingredient of such a renormalization is to take into account not only the wave intensity ($\sim\varepsilon^2$), but also the actual fine structure of intense wave packets \citep{Tao12:GRL, Tao13,Zhang20:grl:frequency, Zhang20:grl:phase}. We can take into account these fine structure effects via the parameter $N_{trap}$, the number of periods of trapped particle motion. Although $N_{trap}$ is not a directly measured wave characteristic, this parameter can include effects of wave-packet shortness \citep{Zhang18:jgr:intensewaves, Zhang20:grl:phase, Mourenas18:jgr, Hiraga&Omura20} and effects of wave coherence destruction \citep{Tao13, Artemyev15:pop:stability, Zhang20:grl:phase}. Alternatively, we can take into account the fine structure of whistler-mode waves via the parameter $\beta$, the wave packet length in number of wave periods, which can be directly measured together with $\varepsilon$ \citep{Zhang18:jgr:intensewaves}. For resonant interactions with $<1$ MeV electrons, we get $D_{NL}/D_{QL}\sim \varepsilon^{-3/2+2\eta/3}\beta^{2\eta/3}$ with a value of $\eta\in[5/2,9/2]$ that depends on assumptions concerning the wave packet distribution in the system. For the rare very long wave packets with $\beta\sim 10^2$ and high wave amplitude $\varepsilon\sim 10^{-3}$ we obtain $D_{NL}/D_{QL}\sim 650$ for $\eta=5/2$ and $D_{NL}/D_{QL}\sim 150$ for $\eta=7/2$, but the time-weighted occurrence rate of such long and intense wave packets varies between $\sim 0.5\times 10^{-3}$ and $\sim 2\times 10^{-3}$ from mildly to strongly active geomagnetic conditions \citep{Zhang18:jgr:intensewaves, Mourenas18:jgr}. In the most realistic case of a long succession of short wave packets (corresponding to $\eta=7/2$) and for wave amplitudes above an approximate threshold $\varepsilon\geq 10^{-3}$ for nonlinear interaction \citep{Omura08,Zhang19:grl}, integrating over $\beta\sim 2-200$ the expression (\ref{eq:25}) of $D_{NL}/D_{QL}$, weighted by the measured occurrence rate of such $\beta$ packets \citep{Mourenas18:jgr}, gives $\langle D_{NL}/D_{QL}\rangle_{\beta}\sim 3$. For smaller amplitudes $\varepsilon<10^{-3}$, representing $\approx 70$\% of all packets \citep{Zhang19:grl}, we should have $D_{NL}/D_{QL}\sim 1$, giving a final approximate ratio $\langle D_{NL}\rangle_{\beta,\varepsilon}/\langle D_{QL}\rangle_{\varepsilon,\beta}\sim 2$. These estimates suggest that over relatively long time periods ($\sim$ days), nonlinear interactions should only slightly speed up the evolution of electron fluxes in comparison with the diffusive evolution for the same time-averaged wave intensity. However, a more accurate parametrical study would be needed to determine for what energies, pitch-angles, and geomagnetic conditions the contribution of nonlinear interactions maximizes. It is also worth noting that the present comparison of relatively long time-scales of electron distribution relaxation does not contradict the fact that only the regime of nonlinear resonant interaction can explain various short-lived events of rapid electron acceleration or loss, which occur much faster than the diffusive evolution \citep[e.g.,][]{Agapitov15:grl:acceleration, Gan20:grl:II}. 

To conclude, we have investigated the transition between two regimes of resonant interaction of electrons and whistler-mode waves in the Earth’s radiation belts: quasi-linear diffusion and nonlinear resonant interaction (including effects of phase trapping and phase bunching). To characterize both regimes within the same framework, we have introduced the typical time-scale $1/D$ of evolution of the electron energy distribution on the resonance curve. This time-scale characterizes the increase of the distribution dispersion for both diffusive ($1/D_{QL}$) and nonlinear ($1/D_{NL}$) relaxations of the electron distribution. For the diffusive relaxation $D_{QL}\sim \varepsilon^{2}$ where $\varepsilon=B_w/B_0$ is the normalized wave amplitude. Thus, we investigated how $D_{NL}$ depends on $\varepsilon$ and on the characteristics of intense wave-packets, which are described by $N_{trap}$, the number of periods in trapped motion, or $\beta$, the wave packet size ($N_{trap}\sim \beta\varepsilon^{1/2}$ when $\beta<1/\varepsilon$) and/or to the closeness of the wave packet amplitude to its threshold value required for nonlinear resonances ($N_{trap}\sim \delta_0^{3/4}\varepsilon^{-1/2}$ with $\delta_0\to 1$ for large amplitude packets). The obtained scaling is $D_{NL}/D_{QL}\sim \varepsilon^{-3/2}\left(\varepsilon \beta\right)^{2\eta/3}$ with $\eta$ varying from $5/2$ (when we compare electron distribution relaxation due to diffusion and nonlinear interactions over the same energy range) to $9/2$ (when diffusion acts over a wide energy range, whereas nonlinear interactions are very rare and occur in a narrower energy range). These scalings can be used for generalizing the existing radiation belt models and for an evaluation of the importance of nonlinear interactions for realistic distributions of wave characteristics.

% If you have acknowledgments, this puts in the proper section head.
\begin{acknowledgments}
The work of A.V.A., A.I.N., and A.A.V. was supported by the Russian Science
Foundation, Project No. 19-12-00313.
%Work of A.I.N. was supported by the Leverhulme Thrust (grant \# RPG-2018-143).
\end{acknowledgments}

% Create the reference section using BibTeX:
%\bibliography{full,addon}
%merlin.mbs apsrev4-1.bst 2010-07-25 4.21a (PWD, AO, DPC) hacked
%Control: key (0)
%Control: author (72) initials jnrlst
%Control: editor formatted (1) identically to author
%Control: production of article title (-1) disabled
%Control: page (0) single
%Control: year (1) truncated
%Control: production of eprint (0) enabled
%

\end{document}